\documentclass[12pt,twoside]{article}
\usepackage{epsfig}
\usepackage{graphicx}
\usepackage{amssymb}
\usepackage{url}
\usepackage{amsfonts,amssymb,amsbsy}
\usepackage{psfrag}
\usepackage{amsmath}
\usepackage{comment}
\title{Soft-and-Hard/D'B' Boundary Conditions\\ and their Realization by Electromagnetic Media}
\author{I.V. Lindell and A.H. Sihvola} 
\date{Department of Radio Science and Engineering\\ Aalto University, School of Electrical Engineering\\ P.O.Box 13000, Espoo 00076AALTO, Finland\\ {\tt ismo.lindell@aalto.fi, ari.sihvola@aalto.fi}}
\def\e{\begin{equation}} 
\def\f{\end{equation}} 
\def\ea{\begin{eqnarray}} 
\def\fa{\end{eqnarray}} 

\def\##1{{\bf #1\mit}}
\def\%#1{{\mbox{\boldmath $#1$}}}
\def\=#1{{\overline{\overline{\mathsf #1}}}}

\def\*{^{\displaystyle*}}

\def\.{\cdot}
\def\x{\times}
\def\:{\over}
\def\oo{\infty}

\def\D{\nabla}
\def\d{\partial}

\def\ra{\rightarrow}

\def\l#1{\label{eq:#1}}
\def\r#1{(\ref{eq:#1})}
\def\am{\left(\begin{array}{c}}
\def\amm{\left(\begin{array}{cc}}
\def\a{\end{array}\right)}

\def\De{\Delta}
\def\E{\epsilon}

\def\h{\eta}

\def\la{\lambda}

\def\M{\mu}
\def\o{\omega}

\def\TH{\theta}

\def\VF{\varphi}

\begin{document}

\maketitle

\begin{abstract}
A layer of uniaxial medium with large axial permittivity and permeability can be used as a quarter-wave transformer with interesting properties. By increasing the transverse permittivity and permeability the transformer becomes a thin sheet. It is shown that the recently introduced SHDB boundary conditions, generalizing the soft-and-hard and DB conditions, realized by the interface of a skewon-axion medium, can be transformed to form a novel class of SHD'B' boundary conditions which generalizes the soft-and-hard and D'B' boundary conditions. Reflection of a plane wave from a planar SHD'B' boundary is considered by numerical examples revealing an interesting narrow beam with radical change of reflection for certain values of parameters and incidence angles.  
\end{abstract}

\section{Introduction}

Various novelties in electromagnetic boundary conditions have been recently introduced and their physical and material realizations have been attempted. For example, the perfect electromagnetic conductor (PEMC) boundary conditions, 
\e \#n\x(\#H+M\#E)=0,\ \ \ \ \#n\.(\#D-M\#B)=0, \l{PEMC}\f
involving one scalar parameter $M$ were introduced in \cite{PEMC,AnnPhys} with material realizations suggested in \cite{PEMC1,Caloz,Attiya}. Here, $\#n$ denotes the unit vector normal to the boundary surface. As another example, the DB-boundary conditions requiring vanishing of the normal components of the $\#D$ and $\#B$ vectors were defined by  \cite{IBDB,DB} 
\e \#n\.\#D=0,\ \ \ \ \#n\.\#B=0. \l{DB}\f
They have found application in cloaking problems \cite{Kong08,Yaghjian,Weder,Kildal09,Martini}. Some physical realizations for the DB conditions in terms of medium interfaces have been suggested in \cite{Rumsey,IBDB,SS} and a more practical material realization was given in \cite{Zaluski}. As a further example, boundary conditions requiring vanishing of normal derivatives of normal field components were defined in \cite{AP10} by
\e \D\.(\#n\#n\.\#D)=0,\ \ \ \ \D\.(\#n\#n\.\#B)=0, \l{D'B'}\f
which for the planar boundary $z=0$ are simplified to
\e \d_zD_z = D_z'=0,\ \ \ \ \ \d_zB_z = B_z'=0. \l{D'B'1}\f
Conditions \r{D'B'} have been dubbed D'B' conditions. While a physical realization in terms of a medium interface has not yet been found, it was shown that the D'B' conditions can be obtained in terms of a layer of a certain medium upon the DB boundary \cite{D'B',Meta}. Generalizations of the DB and D'B' conditions have been discussed in \cite{AWPL09,Mixed}. 

Finally, a novel set of conditions, generalizing those of the soft-and-hard boundary \cite{Kildal} and the DB boundary, was obtained in \cite{SHDB} when considering the interface of a certain class of skewon-axion media \cite{Hehl}, also called IB media \cite{IB}. They were subsequently dubbed SHDB boundary conditions and expressed in the form
\ea A\#n\.\#B + \#u\.\#E&=&0, \l{SHDB1}\\
 A\#n\.\#D -\#u\.\#H&=&0, \l{SHDB2} \fa
where $\#u$ satisfying $\#n\.\#u=0$ is a unit vector parallel to the boundary. For $|A|\ra0$ the conditions can be seen to reduce to those of the soft-and-hard boundary, $\#u\.\#E=0$, $\#u\.\#H=0$, while for $|A|\ra\oo$ they approach the DB conditions \r{DB}. Since all of the recently introduced boundary conditions \r{DB}, \r{D'B'} and \r{SHDB1}, \r{SHDB2} have been shown to be physically realizable in terms of medium interfaces, they cannot be considered just mathematical concepts. However, there still is work to do to go one step further by realizing the medium parameters in terms of metamaterial structures. 

It is the purpose of the present paper to define a set of boundary conditions which is related to the SHDB boundary in the same way as the D'B' boundary is related to the DB boundary, i.e., a boundary which is a generalization of the soft-and-hard and D'B' boundaries. Because the D'B' boundary conditions could be produced by transforming the DB boundary conditions in terms of a layer of uniaxial medium \cite{D'B'}, it appears reasonable to apply a similar idea for the SHDB boundary.

\section{Quarter-wave transformer}

Let us consider a uniaxial anisotropic medium defined by the medium equations
\ea \#D &=& \E_t\#E_t + \E_z\#u_zE_z,\\
    \#B &=& \M_t\#H_t + \M_z\#u_zH_z,\fa
where $\#E_t$ and $\#H_t$ are components transverse to the $z$ axis. Assuming that the axial parameters become infinitely large,
\e \E_z\ra\oo,\ \ \ \ \M_z\ra\oo, \f
we must have $E_z\ra0$ and $H_z\ra0$ in the medium while $D_z$ and $B_z$ may have finite values. In practice it is sufficient that the parameters satisfy
\e \E_z\gg\M_t,\ \ \ \ \ \M_z\gg\M_t. \f
Such a medium has been previously called by the name waveguiding medium \cite{AP06}, because a wave propagates along the axis of the medium like in a set of parallel waveguides. The axial components of the Maxwell equations are
\ea \#u_z\.\D_t\x\#E_t &=& -j\o B_z, \l{Bz}\\
    \#u_z\.\D_t\x\#H_t &=& j\o D_z, \l{Dz}\fa
while the transverse components can be represented in the form 
 \ea   \d_z\#E_t &=& jk_t\h_t\#u_z\x\#H_t, \l{Et}\\
       \h_t\d_z\#H_t &=& -jk_t\#u_z\x\#E_t, \l{Ht}\fa
with
 \e k_t = \o\sqrt{\M_t\E_t},\ \ \ \ \h_t=\sqrt{\M_t/\E_t}. \f
Operating \r{Et} and \r{Ht} by $\d_z$ leads to the equations
\e (\d_z^2+k_t^2)\am \#E_t\\ \#H_t\a = \am 0\\ 0\a. \f
Suppressing the dependence on $x$ and $y$, the transverse fields must be of the form
\e \am \#E_t(z)\\ \#H_t(z)\a = \am \#E_t^+\\ \#H_t^+\a e^{-jk_tz} + \am \#E_t^-\\ \#H_t^-\a e^{jk_tz}. \l{waves}\f 
Actually, $(\#E_t^+, \#H_t^+)$ and $(\#E_t^-, \#H_t^-)$ may have any dependence on $x$ and $y$. Inserting \r{waves} in \r{Et} and \r{Ht} and equating components with similar $z$ dependence, we obtain the following relations for the field amplitude vectors:
\ea \#E_t^+ &=& -\#u_z\x\h_t\#H_t^+ \\
    \#E_t^- &=& \#u_z\x\h_t\#H_t^-. \fa
    
Let us now consider the fields in the waveguiding medium at $z=0$ as related to fields at $z=-d$ when choosing 
\e d = \frac{\pi}{2k_t} = \frac{\la_t}{4},\ \ \ \ \ \exp(\pm jk_td) = \pm j. \f
The relations can be expressed as
\ea   \#E_t(0) &=& j\h_t\#u_z\x\#H_t(-d), \l{Ed}\\ 
      \h_t\#H_t(0) &=& -j\#u_z\x\#E_t(-d). \l{Hd}\fa
The corresponding relations for the axial field components are obtained by substituting \r{waves} in \r{Bz} and \r{Dz}, and the same differentiated by $\d_z$. Skipping the details, the results can be represented as
\ea \am D_z(0)\\ B_z(0)\a &=& \frac{1}{k_t} \am D_z'(-d)\\ B_z'(-d)\a, \l{DBd}\\
    \am D_z'(0)\\ B_z'(0)\a &=& -k_t \am D_z(-d)\\ B_z(-d)\a. \l{D'B'd}\fa

Equations \r{Ed}, \r{Hd}, \r{DBd} and \r{D'B'd} represent relations between fields at the two planes in the waveguiding medium. A layer of quarter wavelength of such a medium transforms boundary conditions at $z=-d$ to other boundary conditions at $z=0$. By letting the transverse parameters grow in magnitude with respect to the parameters $\E,\M$ of the exterior medium, $\E_t/\E\ra\oo$ and $\M_t/\M\ra\oo$, the layer becomes a thin sheet. In such a case there exist two different orders of large parameters, $\E_z\gg\E_t\gg\E$, $\M_t\gg\M_t\gg\M$.

\begin{figure}[htb]
{\centering
   \psfrag{z}[][]{$z$} 
   \psfrag{0}[][]{$-d$} 
   \psfrag{d}[][]{$0$} 
   \psfrag{i}[][]{$i$} 
   \psfrag{e}[][]{$e$} 
   \psfrag{e0}[][]{$\epsilon,\mu$} 
   \psfrag{ki}[][]{${\vec k}^i$} 
   \psfrag{kr}[][]{${\vec k}^r$} 
   \psfrag{kt}[][]{$\E_t,\M_t$} 
   \psfrag{f}[][]{$\displaystyle \frac{\lambda_t}{4}$} 
\includegraphics[width=8cm]{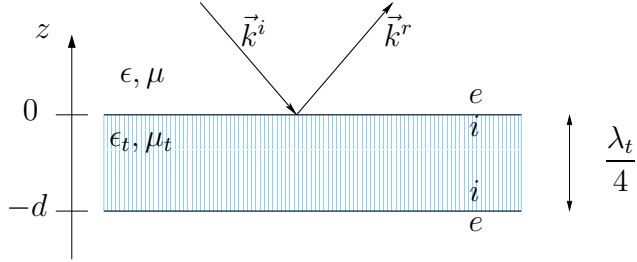}
\caption{\label{fig_layer} Quarter-wave transformer is a layer of uniaxial medium with large axial parameters, $\M_z\gg\M_t$ and $\E_z\gg\E_t$. It can be used for transforming boundary conditions at $z=-d$ to other boundary conditions at $z=0$. For $\M_t\gg\M$ and $\E_t\gg\E$ the layer becomes a thin sheet.}}
\end{figure}

One must note that, while $\#E_t,\#H_t, D_z$ and $B_z$ are continuous through the interfaces, $D_z'$ and $B_z'$ are not. Thus, \r{DBd} and \r{D'B'd} are valid only when $D'_z$ and $B'_z$ are taken at the interior side of the interface. Denoting the interior side of the interfaces by $()_i$ and the exterior side by $()_e$, the relations can be obtained from
\ea \D\.\#B_i &=& \d_zB_{zi} + \M_t\D_t\.\#H_{ti} =0,\\
    \D\.\#D_i &=& \d_zD_{zi} + \E_t\D_t\.\#E_{ti}=0, \\
    \D\.\#B_e &=& \d_zB_{ze} + \M \D_t\.\#H_{te} =0,\\
    \D\.\#D_e &=& \d_zD_{ze} + \E \D_t\.\#E_{te}=0. \fa
Let us assume that the half space $z>0$ is isotropic with parameters $\E,\M$ while the other half space $z<-d$ is taken care of by boundary conditions at $z=-d_e$. Due to continuity of $\D_t\.\#E_t$ and $\D_t\.\#H_t$ through the interface $z=0$, we obtain the relations
\e \M B'_z(0_i)= \M_t B'_z(0_e),\ \ \ \ \E D'_z(0_i)= \E_t D'_z(0_e). \f
Assuming for simplicity $\M_t/\M = \E_t/\E=k_t/k$, with $k=\o\sqrt{\M\E}$, the conditions \r{DBd} and \r{D'B'd} can be written as
\ea \am D_z(0_e)\\ B_z(0_e)\a &=& \frac{1}{k_t} \am D_z'(-d_i)\\ B_z'(-d_i)\a, \l{DBde}\\
    \am D_z'(0_e)\\ B_z'(0_e)\a &=& -k  \am D_z(-d_i)\\ B_z(-d_i)\a. \l{D'B'de}\fa

\section{SHD'B' boundary conditions}

The case of obtaining D'B' conditions \r{D'B'1} by transforming the DB conditions is now obvious from \r{D'B'de}. Also, other cases considered in \cite{D'B'} are directly obtained. Let us now consider the SHDB conditions \r{SHDB1}, \r{SHDB2} at $z=-d_e$, by assuming $\#u=\#u_x$,
\ea A B_z(-d_e) + E_x(-d_e)&=& 0\\
    A D_z(-d_e) - H_x(-d_e)&=& 0. \fa
Since these field quantities are continuous through the interface $z=-d$, we can replace $d_e$ by $d_i$ and apply \r{Ed}, \r{Hd} and \r{D'B'de}. Thus, the SHDB conditions are transformed through the quarter-wave transformer to the form
\ea AB'_z(0_e) -j\o\M H_y(0_e)&=&0, \l{SHD'B'1}\\
 AD'_z(0_e) -j\o\E E_y(0_e)&=&0. \l{SHD'B'2} \fa
Because the boundary conditions \r{SHD'B'1} and \r{SHD'B'2} correspond to a certain physical setup, a layer of waveguiding medium upon a half space of certain medium yielding the SHDB conditions, they have a firm physical significance. It is proposed that they will be called SHD'B' boundary conditions. From the analogy of conditions \r{D'B'} and \r{SHDB1}, \r{SHDB2}, the more general form for the SHD'B' boundary conditions is
\ea A\D\.(\#n\#n\.\#B) + \#v\.(\D\x\#E)&=&0, \l{SHD'B'3}\\
 A\D\.(\#n\#n\.\#D) -\#v\.(\D\x\#H)&=&0, \l{SHD'B'4} \fa
where $\#v=\#n\x\#u$ for $\#u$ defined by  \r{SHDB1}, \r{SHDB2}. In this form the SHD'B' conditions are valid for more generally than for planar boundaries. Making the transformer a thin sheet, and because its transforming property is local, it can be shaped to follow any smooth surface (with radius of curvature assumed much larger than the thickness of the sheet). 

The two extreme values of the parameter $A$ yield two known special cases: $|A|\ra0$ corresponds to the conditions of the soft-and hard boundary while $|A|\ra\oo$ corresponds to those of the D'B' boundary. Since in the form \r{SHD'B'3}, \r{SHD'B'4} the boundary conditions do not involve the frequency, they are valid to any time dependence like the SHDB conditions \r{SHDB1}, \r{SHDB2}. Furthermore, both sets of boundary conditions are self dual. In fact, making the simple duality transformation \cite{Methods} 
\e \#E_d= j\h\#H,\ \ \ \ \ \#H_d=\#E/j\h, \f
with $\h=\sqrt{\M/\E}$, the pair of conditions \r{SHD'B'3}, \r{SHD'B'4} can be seen to remain invariant. Thus, just like the SHDB boundary conditions \r{SHDB1}, \r{SHDB2}, the SHD'B' boundary conditions are self dual.

\section{Plane-wave reflection from SHD'B' plane}

To have an idea of the properties of the SHD'B' boundary, let us consider the basic problem of plane-wave reflection from a SHD'B' plane $z=0$. The incident plane wave in the isotropic half space $z>0$ is defined by
\ea \#E^i(\#r) &=& \#E^i \exp(-j\#k^i\.\#r),\\
    \#H^i(\#r) &=& \#H^i \exp(-j\#k^i\.\#r),\fa
and the field reflected from the SHD'B' is denoted by
\ea \#E^r(\#r) &=& \#E^r \exp(-j\#k^r\.\#r),\\
    \#H^r(\#r) &=& \#H^r \exp(-j\#k^r\.\#r).\fa
The different $\#k$ vectors are   
\ea \#k^i&=&-k_z\#u_z+ \#k_t,\\
 \#k^r&=&k_z\#u_z+ \#k_t,\\
 \#k_t&=& k_x\#u_x + k_y\#u_y. \fa
Applying the Maxwell equations for the fields of the incident and reflected plane waves, 
\ea k\h\#H^i= \#k^i\x\#E^i,\ \ && \ \ k\h\#H^r=\#k^r\x\#E^r, \\
    k\#E^i= -\#k^i\x\h\#H^i,\ \ &&\ \ k\#E^r=-\#k^r\x\h\#H^r, \fa
and the orthogonality conditions
\ea \#k^i\.\#E^i&=&\#k^r\.\#E^r=0, \\
 \#k^i\.\#H^i&=&\#k^r\.\#H^r=0, \fa
relations between the tangential field components can be compactly expressed as
\ea \h\#H_t^i = -\=J\.\#E_t^i,\ \ &&\ \ \h\#H_t^r = \=J\.\#E_t^r, \l{kkH}\\
 \#E_t^i = \=J\.\h\#H_t^i, \ \ &&\ \  \#E_t^r = -\=J\.\h\#H_t^r. \l{kkE}\fa
The 2D dyadic $\=J$ defined by
\e \=J=\frac{1}{kk_z}\#u_z\x(k_z^2\=I_t + \#k_t\#k_t),\f
resembles the imaginary unit because of its properties \cite{IBDB}
\e \=J{}^2 = -\=I_t, \ \ \ \ \=J{}^{-1}=-\=J,\f
which can be easily verified. 

The SHD'B' conditions \r{SHD'B'1}, \r{SHD'B'2} can be expressed in terms of a dimensionless parameter $C$ defined by
\e C=\frac{Ak}{\o}=A\sqrt{\M\E}, \f
in a form involving the tangential field components as
\ea C \d_xH_x(0) +(C\d_y +jk) H_y(0)&=&0, \l{7}\\
    C \d_xE_x(0) +(C\d_y +jk) E_y(0)&=&0. \l{8} \fa
For the plane wave fields, \r{7}, \r{8} become
\ea C k_xH_x(0) +(Ck_y- k) H_y(0)&=& 0, \l{CH}\\
    C k_xE_x(0) +(Dk_y- k) E_y(0)&=&0. \l{CE} \fa
Defining the vector
\e \#c = Ck_x\#u_x +(Ck_y-k)\#u_y= C\#k_t-k\#u_y, \l{c}\f
parallel to the boundary, the conditions \r{CH}, \r{CE} can be expressed for the plane-wave fields in the simple form
\ea \#c\.\#H(0) &=& 0, \l{cH} \\    
    \#c\.\#E(0) &=& 0. \l{cE}\fa
These conditions have an appearance similar to that of the soft-and-hard conditions \cite{Kildal}. However, here the vector $\#c$ depends on the $\#k^i$ vector of the incident wave. 

In terms of the amplitude vectors of the incident and reflected field components, the SHD'B' conditions \r{cH}, \r{cE} have the form
\ea \#c\.(\#H_t^i+ \#H_t^r) &=& 0, \l{cHH}\\    
    \#c\.(\#E_t^i+ \#E_t^r) &=& 0. \l{cEE}\fa    
Applying \r{kkH}, \r{kkE}, the conditions \r{cHH} and \r{cEE} can be respectively rewritten as
\ea \#b\.(\#E_t^i-\#E_t^r)&=&0,\l{bEE}\\
    \#b\.(\#H^i_t-\#H^r_t)&=&0, \l{bHH}\fa 
where $\#b$ is another vector parallel to the boundary,
\e \#b = \#c\.\=J= \frac{1}{kk_z}(k_z^2\#c\x\#u_z- (\#u_z\.\#c\x\#k_t)\#k_t). \l{bc}\f
Conversely, we can write
\e \#c = -\#b\.\=J. \l{cb}\f

In the general case the vectors $\#b$ and $\#c$ are linearly independent, i.e., they satisfy
\e \#b\x\#c\.\#u_z = \frac{1}{kk_z} (\#c\x\#k^i)^2\not=0, \f
whence they make a vector basis in the boundary plane. The vectors $\#b',\#c'$ defined by
\ea \#b' &=& \frac{\#c\x\#u_z}{\#u_z\.\#b\x\#c} ,\\          
    \#c' &=& \frac{\#u_z\x\#b}{\#u_z\.\#b\x\#c} , \fa
form the reciprocal basis satisfying
\e \#b\.\#c'=\#c\.\#b'=0,\ \ \ \ \#b\.\#b'=\#c\.\#c'=1. \f 
The 2D unit dyadic can be expressed as
\e \=I_t = \#b'\#b + \#c'\#c. \l{It}\f
From  \r{bc} and \r{cb} we can write the expansion
\e \=J = \=I_t\.\=J = \#c'\#b -\#b'\#c, \f
whence the reciprocal basis vectors satisfy the relations 
\e \#c'=\=J\.\#b',\ \ \ \ \#b'=-\=J\.\#c'. \f

\section{Eigenfields}

The 2D reflection dyadic $\=R$ can be defined by the relation between the tangential field components as
\e \#E_t^r=  \=R\.\#E_t^i. \l{ERE}\f
Applying \r{cEE}, \r{bEE} and \r{It}, the reflection dyadic in \r{ERE} can be constructed as
\ea \=R &=& \#b'\#b-\#c'\#c. \l{R}\\
 &=& -\frac{\#u_z\x(\#c\#b+\#b\#c)}{\#u_z\.\#c\x\#k_t}. \fa
Defining the angles of incidence $\TH,\VF$ by
\ea k_x&=&k\sin\TH\cos\VF,\\
 k_y&=&k\sin\TH\sin\VF, \\
 k_z &=& k\cos\TH, \fa
the components of the expansion 
\e \=R = \#u_x\#u_x R_{xx} + \#u_x\#u_y R_{xy} + \#u_y\#u_x R_{yx} + \#u_y\#u_y R_{yy}, \f
can be expressed as
\ea R_{xx} &=& \frac{1}{\De}((\cos^2\TH + \sin^2\TH\cos^2\VF)(1-2C\sin\TH\sin\VF) \nonumber\\ &&-C^2\sin^2\TH\cos^2\TH\cos2\VF) = -R_{yy}, \\
   R_{xy} &=& \frac{1}{\De}(\sin^2\TH\sin2\VF(1-C^2\cos^2\TH)\nonumber \\ &&+2C\sin\TH\cos\VF(\cos^2\TH-\sin^2\TH\sin^2\VF)), \\
   R_{yx} &=& \frac{1}{\De}(2C\sin\TH\cos\VF(\cos^2\TH+\sin^2\TH\cos^2\VF) \nonumber\\
   && - C^2\sin^2\TH\cos^2\TH\sin2\VF), \\
   \De &=& \cos^2\VF + (C\sin\TH -\sin\VF)^2\cos^2\TH. \fa
One can easily show that the coefficients $R_{xx}, R_{yy}$ are symmetric, and $R_{xy}, R_{yx}$ are antisymmetric, in the variable $\pi/2 -\VF$. 

Since the eigenproblem for the reflection dyadic \r{R} has the solutions
\e \=R\.\#b' = \#b',\ \ \ \ \=R\.\#c' = -\#c', \f 
the eigenfields corresponding to the eigenvalues $+1$ and $-1$ are respectively
\ea \#E^i_{t+} &=&   E_+\#b'=\#E^r_{t+}, \\
    \#E^i_{t-} &=&  E_-\#c'=-\#E^r_{t-} . \fa
The corresponding magnetic eigenfields are obtained by applying \r{kkH},   \ea \h\#H^i_{t+} &=&  -\=J\.\#E_{t+}^i = -E_+\#c'=-\h\#H^r_{t+} , \\
    \h\#H^i_{t-} &=&  -\=J\.\#E_{t-}^i = E_-\#b'=\h\#H^r_{t-}. \fa
Since the total eigenfields satisfy
\ea (\#E_{-}^i+\#E_{-}^r)_t &=& 0, \\
(\#H_{+}^i+\#H_{+}^r)_t &=& 0, \fa
the $-$ field is reflected as from a PEC plane and, the $+$ field, as from a PMC plane. A corresponding property was previously found for the SHDB boundary \cite{SHDB}.

Applying again $\#k^i\.\#E^i_\pm=0$ and $\#k^i\.\#H_\pm^i=0$, expressions for the total eigenfields can be reduced to
\ea \#E_+^i &=&  E_+\frac{\#c\x\#k^i}{\#k^i\.\#b\x\#c}, \l{E+i}\\
\#E_-^i &=&  E_-\frac{\#k^i\x\#b}{\#k^i\.\#b\x\#c}, \l{E-i}\\
\h\#H_+^i &=&  E_+\frac{\#b\x\#k^i}{\#k^i\.\#b\x\#c}, \l{H+i}\\
\h\#H_-^i &=&  E_-\frac{\#c\x\#k^i}{\#k^i\.\#b\x\#c}. \l{H-i}\fa
The field amplitudes are obtained from the total field as
\e E_+ = \#b\.\#E^i,\ \ \ \ E_-=\#c\.\#E^i. \f

\section{Special cases}

Let us consider some special cases of the SHD'B' medium and the incident plane wave. 
 
\subsection*{Soft-and-hard boundary}

For the case $C=0$ the SHD'B' boundary reduces to the soft-and-hard boundary with boundary conditions
\e \#u_y\.\#E=0,\ \ \ \ \#u_y\.\#H=0. \f
Because \r{c} yields $\#c=-k\#u_y$, and from \r{E+i}, \r{H-i} both $\#E_+^i$ and $\#H_-^i$ have the polarization $\#u_y\x\#k^i$, they satisfy 
\e \#u_y\.\#E_+^i=0,\ \ \ \#u_y\.\#H_-^i=0. \f 
Thus, the $+$ and $-$ waves can be respectively called TE$_y$ and TM$_y$ waves.  The reflection dyadic is simplified to 
\e \=R = \#u_x\#u_x -\#u_y\#u_y +\frac{\sin^2\TH\sin2\VF}{\sin^2\TH\cos^2\VF+ \cos^2\TH}\#u_x\#u_y. \l{RSH}\f

\subsection*{D'B' boundary} 

For $C\ra\oo$ we have $\#c \ra C\#k_t$, whence both $\#E_+^i$ and $\#H_-^i$ have the polarization $\#k_t\x\#k^i= k_z\#k_t\x\#u_z$. In this case the eigenfields satisfy 
\e \#u_z\.\#E_+^i=0,\ \ \ \#u_z\.\#H_-^i=0. \f
The $+$ and $-$ waves can now be respectively called TE$_z$ and TM$_z$ waves. In this and the previous case field decomposition in TE and TM components is independent of the $\#k^i$ vector, whence it is valid for any fields. The reflection dyadic can be written in this case as
\e \=R =  -\cos2\VF(\#u_x\#u_x-\#u_y\#u_y) - \sin2\VF(\#u_x\#u_y+\#u_y\#u_x).\l{RD'B'}\f
         
\subsection*{Special angles of incidence}

Let us consider special cases of the vector $\#k^i$.

\begin{itemize}
\item In the case $\#u_x\.\#k^i=k_x=0$, or $\VF=\pi/2$ the reflection dyadic \r{R} has the simple form
\e \=R = \#u_x\#u_x - \#u_y\#u_y. \f
For a wave with this incidence the boundary appears as a hard surface \cite{Kildal}.
\item The same result is obtained for the case $\#k_t=0$, or $\TH=0$.
\item In the case $\#u_y\.\#k^i=k_y=0$, or $\VF=0$, the reflection dyadic takes the form,
\ea \=R &=& \frac{1-C^2\sin^2\TH\cos^2\TH}{1+ C^2\sin^2\TH\cos^2\TH} (\#u_x\#u_x-\#u_y\#u_y) \nonumber\\
&+& \frac{2C\sin\TH(\cos^2\TH\#u_x\#u_y+ \#u_y\#u_x)}{1+ C^2\sin^2\TH\cos^2\TH}. \fa
This expression coincides with the ones above for $C=0$ and $C\ra\oo$.
\end{itemize}

\section{Numerical examples}

\begin{figure}[htb]
\centering{\includegraphics[width=9cm]{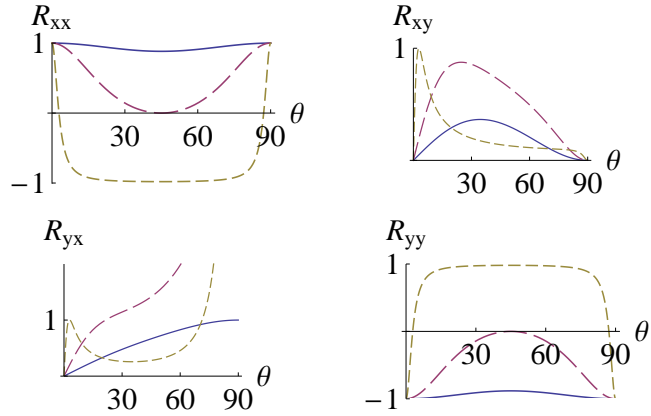}
  \caption{\label{fig:P1} The four components of the reflection matrix
    as function of the incidence angle $\TH$ with three values for the
    parameter $C$: solid line ($C=0.5$), long-dashed line ($C=2$),
    short-dashed line ($C=20$). The azimuth angle is $\VF=0$.}}
\end{figure}

Let us consider as representative examples the reflection coefficients $R_\mathrm{ij}$ for some incident plane waves and some values of the SHD'B'
  boundary parameter $C$. Figure~\ref{fig:P1} illustrates the four reflections coefficients as functions of the incidence angle $\TH$ when the
  plane of incidence is the $xz$ plane ($\VF=0$). The figure shows clearly that $R_{xx}=-R_{yy}$ for all parameter values of $\TH$ and $C$. Another interesting observation is that for $C=2$ and $\TH=45^\circ$, the co-polarized reflection vanishes. In other words the surface acts as a twist polarizer.

\begin{figure}[htb]
\centering{\includegraphics[width=8cm]{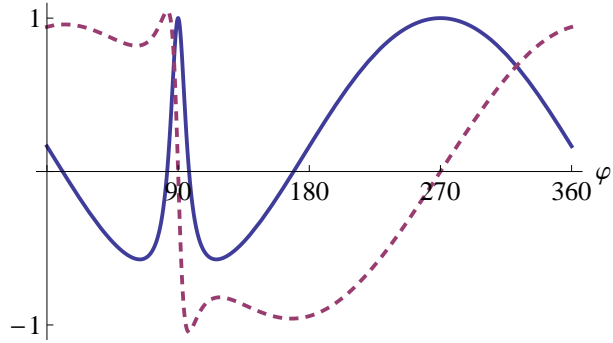}
  \caption{\label{fig:P2} The components $R_\mathrm{xx}$ (solid line)
    and $R_\mathrm{xy}$ (dashed line) of the reflection matrix
    \r{ERE} as function of the azimuth angle $\VF$ for incidence
    angle $\TH=0.3$ and the parameter $C=3$.}}
\end{figure}

Due to the two-dimensional anisotropy of the surface, the reflection characteristics should depends on the azimuthal angle $\VF$. This is
displayed in Figure~\ref{fig:P2} where the co-polarized component $R_\mathrm{xx}(\VF)$ and the cross-polarized component
$R_\mathrm{xy}(\VF)$ are depicted for $\TH=0.3$ and $C=3$. The curves show that $R_\mathrm{xx}$ is symmetric with respect to the incidence
$\pm y$-axis ($\VF=90^\circ$ and $\VF=270^\circ$) whereas $R_\mathrm{xy}$ is antisymmetric. Similarly, $R_\mathrm{yy}$ turns out to be symmetric and $R_\mathrm{yx}$ antisymmetric. The effect of the narrow-beamed radical change in reflection occurs when both $\cos\VF$ and $C\sin\TH-1$ have small values. In such a case the reflection coefficient can be approximated by
\e R_{xx}\approx \frac{(C\sin\TH-1)^2\cos^2\TH - \cos^2\VF}{(C\sin\TH-1)^2\cos^2\TH + \cos^2\VF}. \f
At $\cos\VF=0$ we have $R_{xx}=1$ (exactly) while the nulls of $R_{xx}$ appear at $\cos\VF\approx \pm(C\sin\TH-1)\cos\TH$.

\begin{figure}[htb]
\centering{\includegraphics[width=8cm]{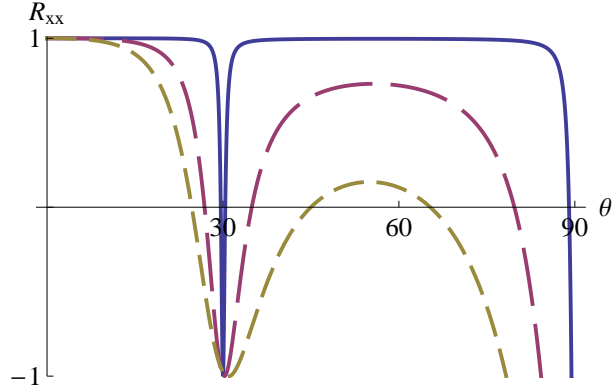}
  \caption{\label{fig:P3} $R_\mathrm{xx}$ as function of the incidence
    angle $\TH$ when the plane of incidence is close to $yz$ plane
    ($\VF=\pi/2$). The azimuth angle is $\VF=\pi/2-0.01$ (solid line),
    $\VF=\pi/2-0.1$ (long-dashed line), and $\VF=\pi/2-0.2$
    (short-dashed line). The boundary parameter has the value $C=2$.}}
\end{figure}

Figure~\ref{fig:P3} shows how the reflection approaches that of a hard surface when the incidence plane comes closer to $yz$ plane, $\VF\ra\pi/2$. For the hard surface has $R_\mathrm{xx}=+1$ independently of the incidence angle, as can be seen from the figure. Again, an interesting narrow-beamed change of reflection at $\TH=\mathrm{arcsin}(1/C)\approx 0.524 = 30^\circ$ can be observed. 

\section{Conclusion}

A novel set of electromagnetic boundary conditions was defined in the present paper. Because the soft-and-hard conditions and the D'B' conditions appear as special cases of the boundary conditions, the generalization was dubbed SHD'B' conditions. The present paper is a continuation to a previous one \cite{SHDB} generalizing the soft-and-hard and DB conditions to what were called the SHDB conditions. Since the D'B' conditions can be realized in terms of a medium layer transforming the DB conditions, the same idea was applied to the realization of the SHD'B' conditions. Making the layer a thin sheet by proper choice of the parameters makes it possible to apply the transformation for the SHDB boundary of any form. To see basic properties of the boundary, reflection of a plane wave of a planar SHD'B' boundary was considered. Analytic expressions for the reflection dyadic and its eigenfields were derived. It was shown that for the two eigenpolarizations the SHD'B' boundary appears as the PEC or the PMC boundary with the respective reflection coefficients $+1$ and $-1$. Numerical computations have shown that,  for certain values of the parameter $C$ and angle of incidence, the field reflected from the SHD'B' boundary is totally cross polarized. Also, there exists an interesting narrow-beam region for the incident wave where the reflection makes a radical change in magnitude. Such a propery could have an application similar to the corner reflector in defining a given direction in space. This may serve as a reason to study the properties further and attempt to find a metamaterial realization for the SHD'B' boundary.

\end{document}